\RequirePackage[2020-02-02]{latexrelease}
\documentclass[%
 reprint, 
 amsmath,amssymb,
 aps,
 nofootinbib
 prb,
 superscriptaddress
 ]{revtex4-1}
\usepackage{scrextend}
\usepackage{graphicx}
\usepackage{dcolumn}
\usepackage{bm}
\usepackage{tabularx}
\usepackage[utf8]{inputenc}
\usepackage[english]{babel}
\usepackage{color}

\usepackage{physics}

\DeclareMathAlphabet{\pazocal}{OMS}{zplm}{m}{n}

\usepackage{hyperref}

\begin{document}

\preprint{APS/123-QED}

\title{Emergence of half-metallic ferromagnetism in transition metal 
substituted Na$_{0.5}$Bi$_{0.5}$TiO$_3$}

\author{Chandan Kumar Vishwakarma}
\affiliation{Department of Physics, Indian Institute of Technology,
             Hauz Khas, New Delhi 110016, India}
\author{B. K. Mani}
\email{bkmani@physics.iitd.ac.in}
\affiliation{Department of Physics, Indian Institute of Technology,
Hauz Khas, New Delhi 110016, India}

\date{\today}

             
\begin{abstract} 

The multifunctional materials with prominent properties such as electrical, 
ferroelectric, magnetic, optical and magneto-optical are of keen interest to 
several practical implications. In the roadmap of designing such materials, 
in the present work, using density functional theory based first-principles 
calculations, we have investigated the functional properties of transition metal 
substituted-NBT. Our calculations predict the emergence of half-metallic 
ferromagnetism in the system. A nonzero magnetic moment of 1.49 
$\mu_{\rm B}/{\rm f.u.}$ is obtained for 25\% concentration of Ni. 
Our data on optical properties for pure NBT is in excellent agreement with
available theory and experiments. For Ni-NBT, we observed a diverging
nature of static dielectric constant, which could be attributed to 
the induced metallic character in the material.  Our simulations on MOKE 
predict a significant Kerr signal of 0.7$^\circ$ for 6.25\% Ni-concentration. 

\end{abstract}

\maketitle

\section{Introduction}

The development of multifunctional materials with two or more properties, such 
as magnetic, ferroelectric, piezoelectric, and optical, has received a lot of 
interest in recent years \cite{Peng-2014, Wang-2005, Zheng-2013}. These materials 
have the potential to revolutionize various industry applications, including 
healthcare, energy and electronics \cite{Lee-2020,Zheng-2021,Huang-2020}. 
In the search for such materials, sodium bismuth titanate, 
Na$_{0.5}$Bi$_{0.5}$TiO$_3$ (NBT), has received a 
remarkable attention than any other lead-free ferroelectrics due to its 
tendency to show multifunctionality by various 
mechanisms \cite{Smolensky-1985, Jones-2002, Schmidt-1995}. 
NBT is a complex perovskite oxide with two cations (Na$^+$ and Bi$^{3+}$) on the A-site 
and one cation (Ti$^{2+}$) on B-site with a rhombohedral symmetry at room 
temperature \cite{smolenskii-1960}. 
It exhibits various anomalous properties associated with site-specific substitutions, 
including improved ferroelectricity and piezoelectricity, magnetism and optoelectronic 
properties \cite{Mishra-2019,Zannen-2015}.

The presence of Ti at the B site provides a strategy to introduce ferromagnetism 
by substituting transition-metal (TM) at B-site. In experimental 
studies, Refs. \cite{Jain-2020} and \cite{Wang-2009}, ferromagnetism at room 
temperature was observed for Fe and Co-doped NBT, respectively. 
In a similar experimental work by Dung {\em et al.,} a room-temperature 
ferromagnetism was reported for Ni-doped NBT \cite{Dung-2020}. 
The maximum magnetization value reported was around $0.91$ $\mu_{\rm B}$/Ni 
for 9\% of Ni concentration at 5 K. Moreover, it was also observed in the 
same study that the optical bandgap decreases with Ni-concentration. 
However, in a different experimental study, Pradhan {\em et al.,} 
the optical band gap was observed to increase with Ni 
concentrations \cite{Pradhan-2018}.
The contradictory trend of experimental data suggest the lack of understanding 
for optical behavior of TM-doped NBT. In addition, to the best of our knowledge, 
there are no data from theory simulations on probing magnetism in TM-doped NBT.
It can thus be surmised that there is a need for a systematic theoretical 
study to understand the underlying mechanism behind the multifunctional 
properties in TM-doped NBT.

The present study aims to probe, with the help of the state-of-the-art of 
first-principles calculations, the electronic, magnetic, optical and 
magneto-optical properties of NBT and TM-substituted NBT. More precisely, 
we aim is to address the following questions:
i) What is the impact of Ni substitution on the electrical and optical properties of NBT?  
ii) Assimilate the mechanism behind the advent of magnetic degrees of freedom in Ni-substituted NBT 
iii) How this introduced ferromagnetism couples with the dielectric properties of NBT?
To assess the coupling between magnetic and optical degrees of freedom, we have 
examined the linear magneto-optic Kerr effect in the polar geometry, in which the 
spin and incident photons are perpendicular to the sample surface. 
This configuration of the Kerr effect is the most favorable way to trace 
the magneto-optical properties experimentally \cite{Banstjan-2008,Ramesh-2009,Malakhovskii-2012}.

The texts in the paper are organized in four sections. In Section II, we provide 
a brief description of the computational methods used in our calculations. 
In Section III, we present and analyze our results on electronic structure, magnetic, 
optical, and magneto-optical properties for NBT and Ni-substituted NBT. 
The summary of our findings is presented in the last Section.

\section{Computational Methodology}

Probing transition metal-substituted NBT structure and emerging 
properties requires an accurate treatment of interstitial effects 
in the material at atomic scale. 
For this, we have performed \textit{ab-initio} spin-polarized calculations 
using density functional theory (DFT) as implemented in the Vienna \textit{ab-initio} 
simulation package (VASP)\cite{kresse-1996,kresse-1996PRB}. To account for the exchange
correlation among electrons, we used Perdew-Burke-Ernzerhof (PBE) \cite{PBE96} 
variation of generalized-gradient approximation pseudopotential. 
And, to account for the strongly correlated $3d$-electrons of Ni we have
incorporated the Hubbard U correction \cite{vladimir91} in our calculation. 
The value, 11.57,  of U is computed self-consistently using density 
functional perturbation theory (DFPT) employing 
the cococcioni's \textit{et al.} \cite{cococcioni05} approach. 
A rhombohedral supercell of size $2\times2\times2$ with 80-atoms is 
used to incorporate various concentrations of Ni.  
All the structures were optimized using full relaxation calculations
up to $10^{-4}$ eV \AA$^{-1}$ force tolerance. For this, we used conjugate 
gradient algorithm with Monkhorst-Pack \cite{monkhorst76} $k$-mesh 
of {\bf $5\times5\times5$}. For the self-consistent-field (SCF) calculations, 
the Brillouin zone was sampled with {\bf $9\times9\times9$} $k$-mesh. 
The energy convergence criterion is maintained at $0.001$ meV, whereas 
the plane wave energy cutoff used was $600$ eV. The real and imaginary 
parts of the dielectric function is calculated using DFPT as 
implemented in VASP.

\section{Results and Discussions}

\begin{figure}
	\centering\includegraphics[scale=0.38, angle=0]{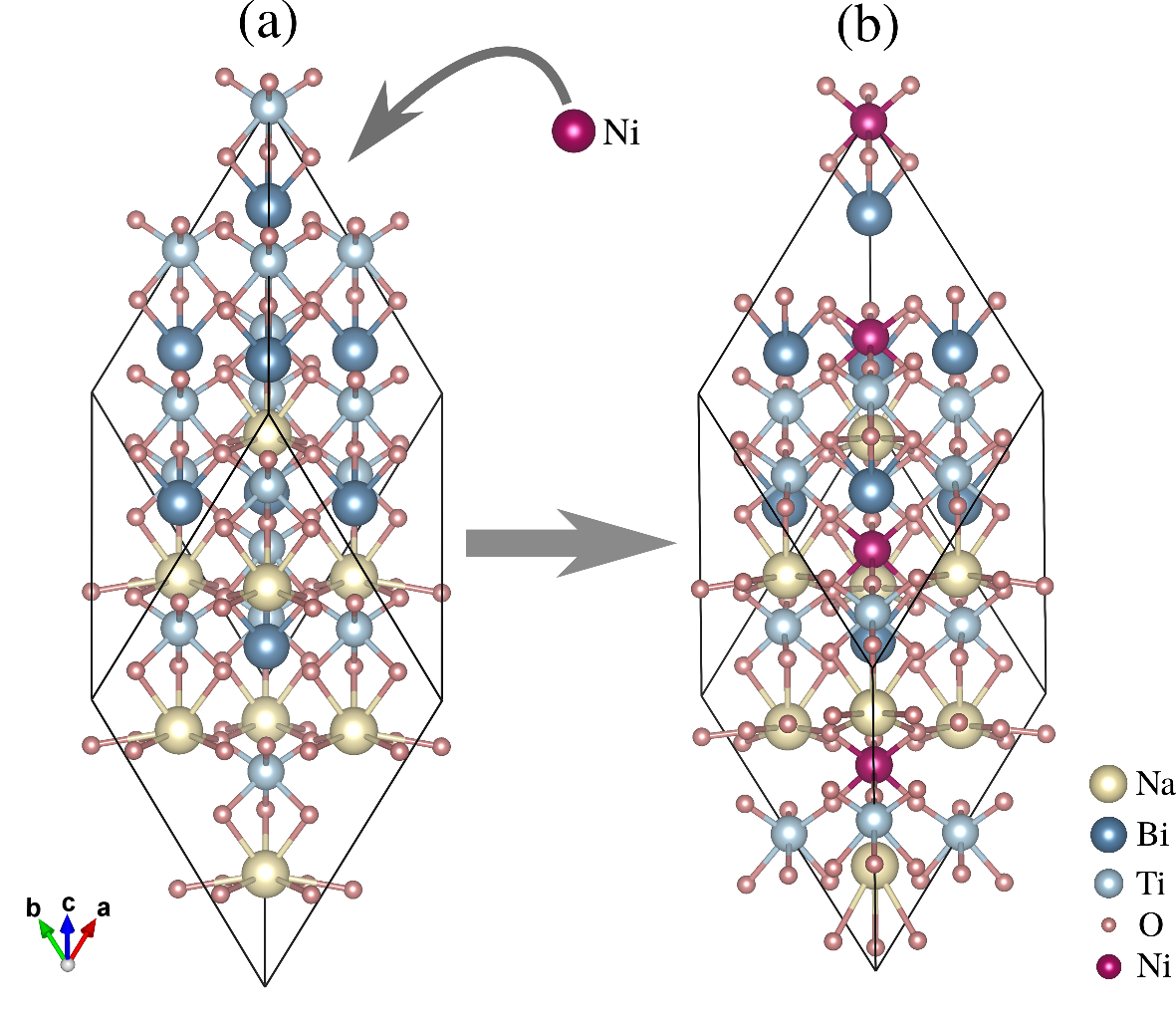}
	\caption{Crystal structures of Na$_{0.5}$Bi$_{0.5}$TiO$_3$ (panel (a)) 
	        and Na$_{0.5}$Bi$_{0.5}$(Ti$_{0.75}$Ni$_{0.25}$)O$_3$ 
		(panel (b)).}
	\label{crystal}
\end{figure}

\subsection{Crystal Structure}

The structural parameters for pure NBT were taken from the experimental crystal 
structure (space group $ R3c $) data \cite{Jones-2002} and optimized further through 
the full relaxation calculations to achieve a minimum energy configuration. Our 
computed lattice parameters and Wyckoff positions are given in Table \ref{tab-lat},
along with the data from the literature for comparison \cite{manal-2022} 
As we observed from the table, our computed lattice parameter $5.65$ is 
in good agreement with the experimental value 5.51 \cite{Jones-2002}.  
The reason for slightly larger value could be attributed to the use of 
GGA functional our calculation \cite{Stampfl-1999}. 

\begin{figure}
	\centering\includegraphics[scale=0.48, angle=-90]{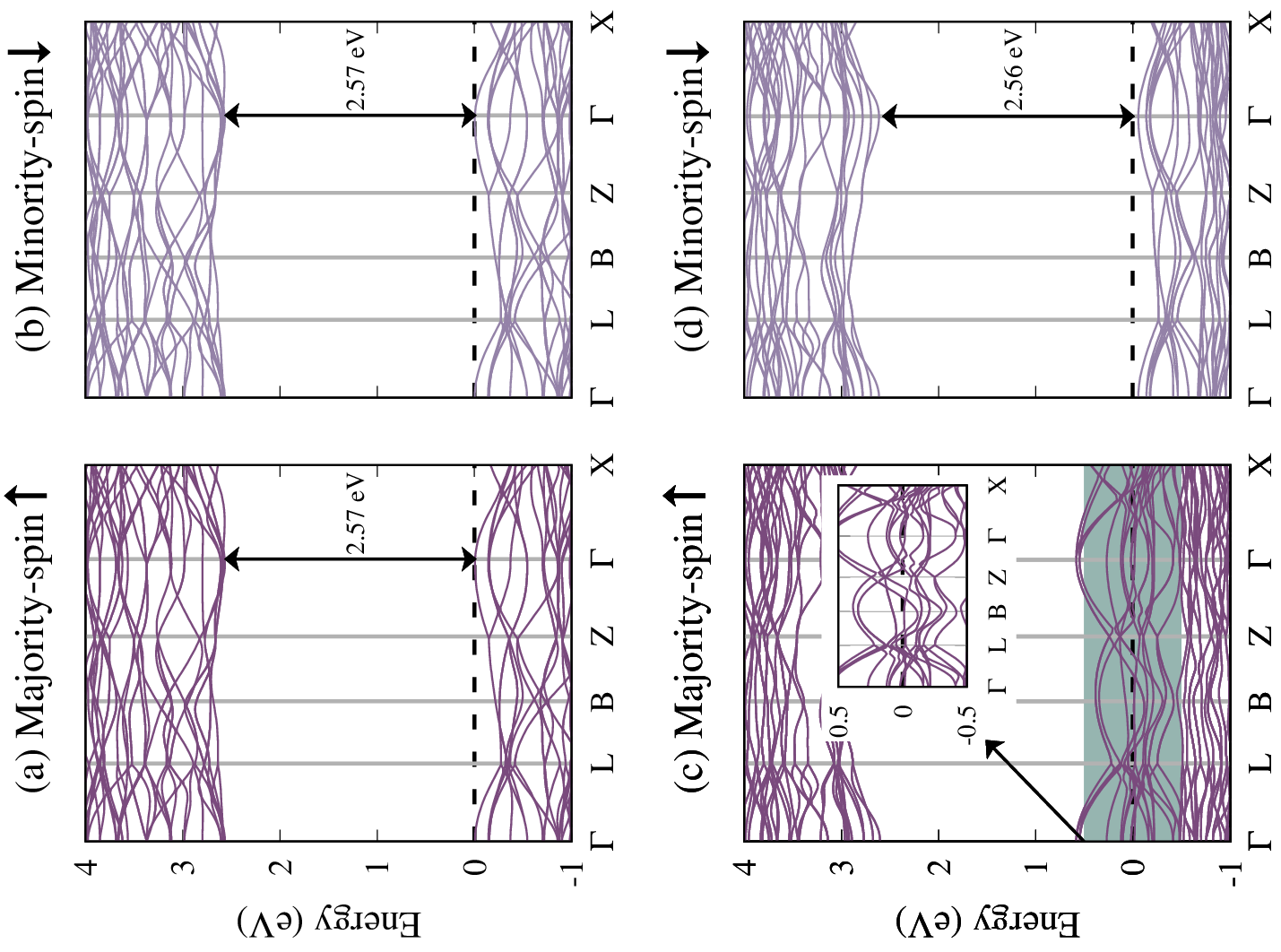}
	\caption{Calculated electronic band structure of Na$_{0.5}$Bi$_{0.5}$TiO$_3$ 
		(panels (a), (b)) and Na$_{0.5}$Bi$_{0.5}$(Ti$_{0.75}$Ni$_{0.25}$)O$_3$ 
		(panels (c), (d)) for majority and minority spin electrons.}
	\label{cryst}
\end{figure}

To incorporate various Ni concentrations in Na$_{0.5}$Bi$_{0.5}$[Ti$_{1-x}$Ni$_x$]O$_3$
(Ni-NBT), we used optimized NBT structure and created a $2\times2\times2$ supercell. And, we 
investigated the properties for x $= 0.0625, 0.125, 
0.1875$, and $0.25$ concentrations of Ni. 
The Ni-NBT structures were fully optimized again using the force tolerance 
up to the $10^{-5}$ eV \AA$^{-1}$. 
From our simulations we find that all Ni-NBT structures crystallize in 
rhombohedral (R$3$m) phase. In Fig. \ref{cryst}, we have shown the crystal 
structures for NBT (panel(a)) and 0.25Ni-NBT (panel(b)).  
The optimized lattice parameters for the chosen concentrations of Ni are 
given in Table \ref{tab-lat}. To the best of our knowledge, there are no 
experimental or other theory data for lattice parameters for Ni-NBT available 
in the literature for comparison.

\begin{table}
	\caption{ Computed lattice parameters and Wyckoff positions for NBT.  
	Values given in the parenthesis are the data from the 
	experiment \cite{manal-2022}.}
\centering
\begin{ruledtabular}
 \begin{tabular}{cccc}
	 &   Lattice Parameters & &\\ 
	 & $a = b = c = 5.647 (5.51)$ (\AA) &  & \\
	 & $\alpha = \beta = \gamma = 58.758 (59.803)(^{\circ})$ & & \\     
	\hline  
	 &   Wyckoff Positions & &\\ 
	Na1 (1a)  & 0.25644 & 0.25644 & 0.25644 \\
	
	Bi1 (1a)  & 0.77324 & 0.77324 & 0.77324 \\
	
	Ti1 (1a)  & -0.00056 & -0.00056 & -0.00056 \\
	
	Ti2 (1a)  & 0.48900 & 0.48900 & 0.48900 \\
	
	O1 (3b)   & 0.23013 & 0.72562 & -0.76987 \\
	
	O2 (3b)   & 0.72771 & 0.21016 & -0.27229 \\
\end{tabular}
\end{ruledtabular}
\label{tab-lat}
\end{table}


\subsection{Electronic Structure and Ferroelectric Properties}

In Fig. \ref{cryst} we have shown the spin-polarized electronic band 
structures of NBT (panels (a) and (b)) and 25Ni-NBT crystals (panels (c) 
and (d)). We chose to report the data for the highest concentration of Ni as 
it has the largest effect on the computed properties. The corresponding data 
for other concentrations are, however, provided in the supplementary material.
As we observed from the panels (a) and (b) of the figure, NBT exhibits a 
direct band gap electronic structure at $\Gamma$ point for both the spin channels.
The calculated band gap, 2.57 eV, is in good agreement with the other 
theoretical value, 2.82, reported in \cite{hongfeng14}.
The observed wide band gaps for both spin channels suggests the semiconducting 
nature of the NBT crystal, and is consistent with the data reported in the 
literature \cite{manal-2022, Zeng-2010, Behara-2020}.
For 25Ni-NBT, however, we observe an asymmetry in the majority and 
minority spin channels (panels (c) and (d)). For majority spin, the Fermi level 
lies in the valence band and shows a metallic nature. Whereas, for minority 
spin sub band, a large band gap of $\sim 2.56$ eV is obtained, which resembles 
the electronic structure of NBT shown in panel (b). This mixed nature of 
electronic structure indicates a half-metallic character of 25Ni-NBT. A similar 
electronic structures we also obtain for other concentrations of Ni.

\begin{figure}
\centering\includegraphics[scale=0.8, angle=0]{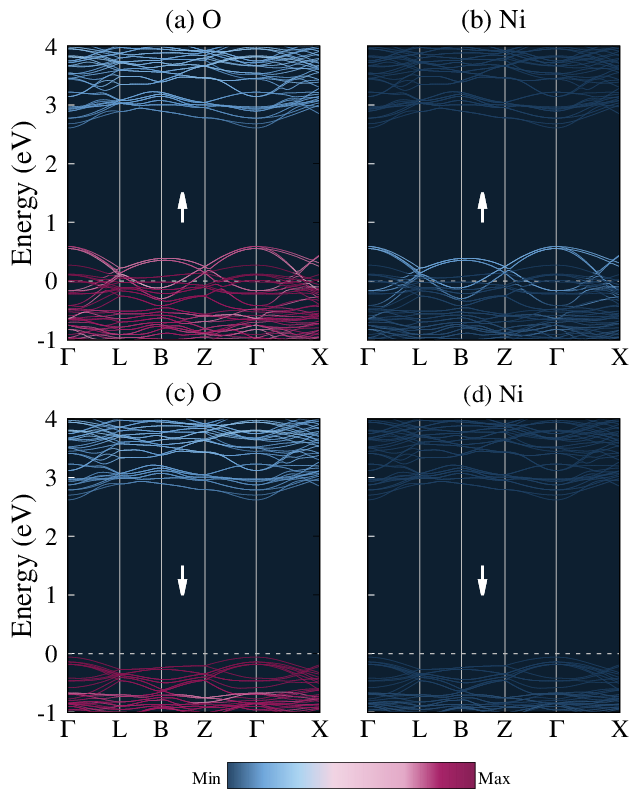}
\caption{The atom-projected band structure of 
	Na$_{0.5}$Bi$_{0.5}$(Ti$_{0.75}$Ni$_{0.25}$)O$_3$ for O and Ni atoms 
	for majority (panels (a), (b)) and  minority (panels (c), (d)) spin 
	electrons.}
\label{fig3}
\end{figure}

\begin{figure}
	\centering\includegraphics[scale=0.35, angle=-90]{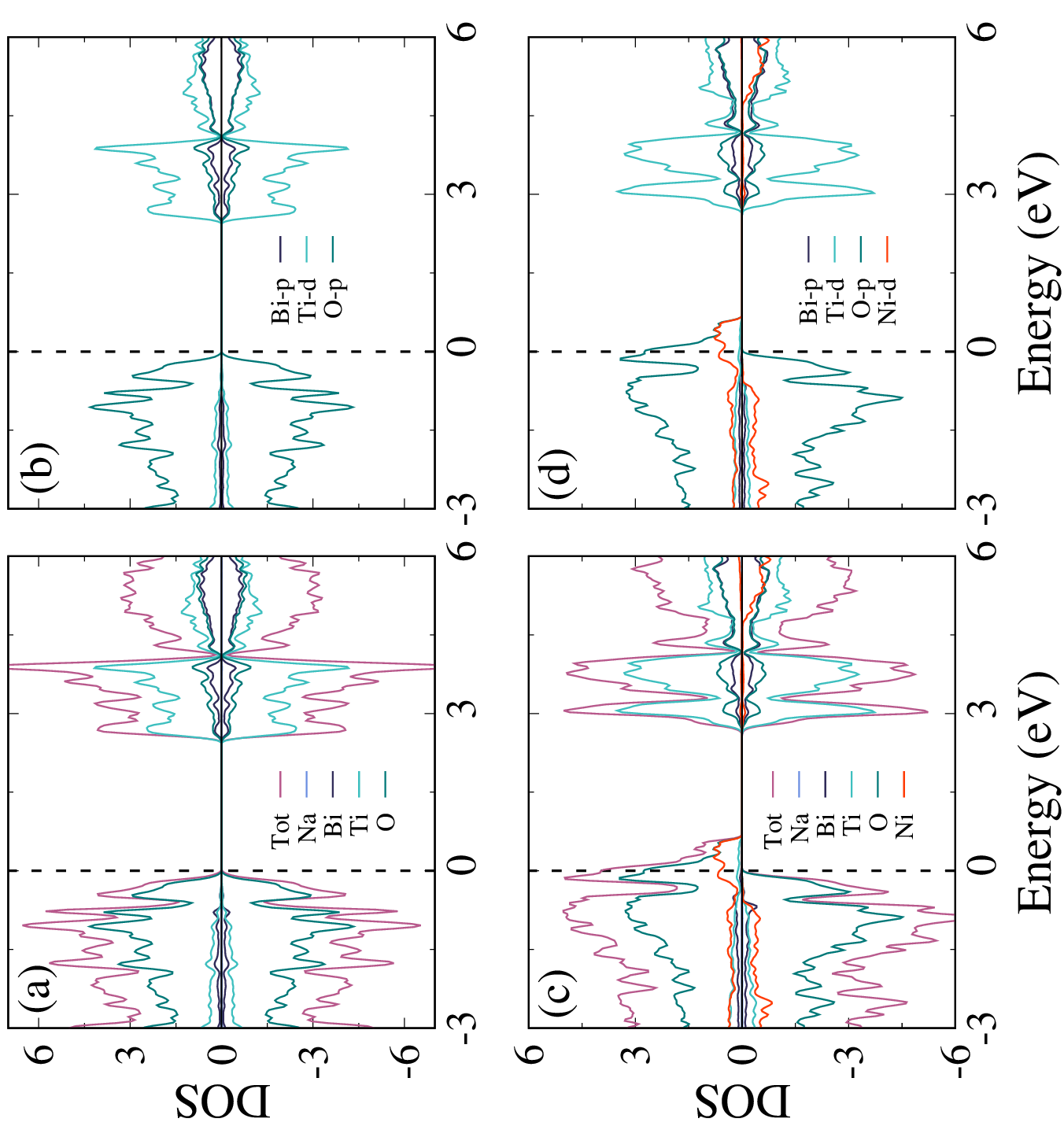}
	\caption{The density of states and orbital-projected density of states for 
		Na$_{0.5}$Bi$_{0.5}$TiO$_3$ (panels (a) and (b)) and 
		Na$_{0.5}$Bi$_{0.5}$(Ti$_{0.75}$Ni$_{0.25}$)O$_3$ 
		(panels (c) and (d)).}
	\label{fig4}
\end{figure}

To get further insight into the half-metallicity in Ni-NBT, we examined 
the atom-projected and orbital-projected electronic structures of NBT and 
25Ni-NBT. The data from this for bands and density of states (DoS) are shown 
in Figs. \ref{fig3} and \ref{fig4}, respectively.
For NBT, as discernible from the panels (a) and (b) of Fig. \ref{fig4}, the
valence band for both the spin channels have dominant contributions from O, 
where $2p$-electrons contribute the most. For the conduction band, however, 
the most significant contribution comes from the $3d$-electrons of Ti.
This observed nature of the electronic structure of NBT is consistent 
with the reported trend in Refs \cite{Zeng-2010}.

For 25Ni-NBT, for majority spin channel, the bands around the Fermi 
energy are of mixed O and Ni character, with O contribution more prominent 
than Ni at the Fermi energy (panels (a), (c) in Fig. \ref{fig3}). This is 
also consistent with the atom-projected DoS shown in Fig \ref{fig4}(b). 
At Fermi energy, O contributes $\approx$ 70\% which mainly comes from 
$2p$-electron, whereas the contribution from Ni (mostly from $3d$-electrons) 
is about $\approx$ 20\% of the total value. And like NBT, the 
conduction band is dominated by the $3d$-electrons of Ti.
For minority spin sub band, a significant contribution of O in the valance 
band below the Fermi level is observed, and comes from the $2p$-electrons.
However, unlike majority spin sub band, there is a negligible contribution 
from Ni in the bands below the Fermi level. Using the number of electronic 
states at Fermi level for both spin 
channels we calculated the spin polarization, which comes out to be 100\%.
This non-zero electronic states at the Fermi level for majority spin and a 
wide band gap for minority spin confirms the half-metallicity in 
Ni-substituted NBT. A similar trend of half-metallicity and spin polarization 
is also observed for Fe-NBT.

Next, we examine the ferroelectric properties in NBT and Ni-NBT. For this, we 
calculated the remanent polarization for NBT and Ni-NBT. The electronic contribution
to polarization was computed using the Berry phase 
approach \cite{King-1993, Resta-1994}. NBT is a well known lead-free ferroelectric 
material with experimentally reported remanent polarization as 
38 $\mu {\rm C}/{\rm cm}^2$ \cite{smolenskii-1960}, 
32 $\mu {\rm C}/{\rm cm}^2$ \cite{Yu-2008} and 
42.4 $\mu {\rm C}/{\rm cm}^2$ \cite{Hari-2017} along [001] pseudo-cubic 
direction. From theory calculations, the reported value of spontaneous 
polarization, $P_{\rm S}$, is 26 $\mu {\rm C}/{\rm cm}^2$ \cite{Niranjan-2013}. 
Our computed value of $P_{\rm S}$ is 49 $\mu {\rm C}/{\rm cm}^2$. The reason for the
difference from experiment could be attributed to the fact that the 
reported experimental polarizations are at room temperature, whereas the
computed values are at 0 K. 
The spontaneous polarization deceases after the Ni-substitution. We obtained 
33.7 and 29.7 $\mu {\rm C}/{\rm cm}^2$ of $P_{\rm S}$ for 12 and 25\% of Ni, respectively.
The reason for this decease could be attributed to the increasing metallic 
nature due to Ni-substitution.



\subsection{Optical Properties}

\begin{figure*}
	\centering\includegraphics[scale=0.5, angle=-90]{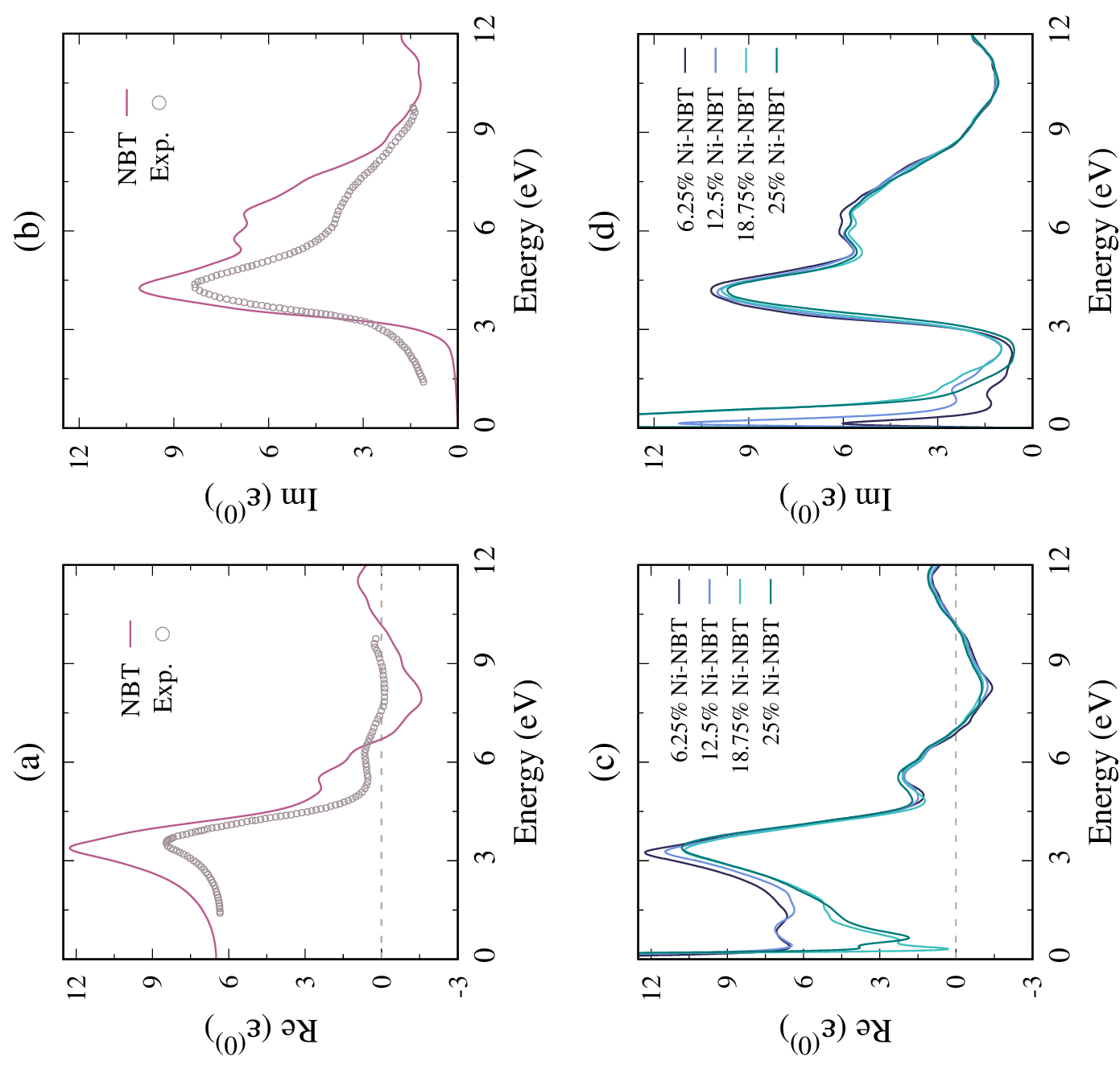}
	\caption{The real and imaginary components of the diagonal permittivity of 
		NBT (panels (a) and (b), respectively)  and Ni-substituted NBT 
		(panels (c) and (d), respectively). The experimental data in panels (a) 
		and (b) are from the Ref. \cite{Bohdan-2009}.}
	\label{opt}
\end{figure*}

Next, we investigate the optical properties of NBT and Ni-NBT. For this, we 
calculated the frequency-dependent complex dielectric function using Ehrenreich 
and Cohen's equation \cite{Ehrenreich-1959}, $\epsilon(\omega) = \epsilon_{1}(\omega) 
+ i\epsilon_{2}(\omega)$, where $\epsilon_{1}$ and $\epsilon_{2}$ are the real 
and imaginary parts, respectively.
The complex dielectric function of a material is a key parameter and could be 
useful in probing several fundamental properties of the material. The imaginary 
part of the dielectric function could be calculated using the linear 
response theory \cite{gajdo06} as 
\begin{equation}
\epsilon_2(\omega) = \frac{2\pi e^2}{\epsilon_0 \Omega} \sum_{k, v, c} 
\delta(E_k^c - E_k^{v} - \hbar\omega) \; 
|\bra{\Psi_k^c}({\mathbf{\hat{n}.r}})\ket{\Psi_k^{v}}|^2.
\end{equation}
Here, the indices $k$, $v$ and $c$ represent the wave vector, valence 
and conduction bands, respectively. The states $|\Psi_k^v \rangle$ and 
$|\Psi_k^c\rangle$ are the wavefunctions associated with valence 
and conduction bands, respectively, and, $E_k^v$ and $E_k^c$ are the 
corresponding energies. The constants, $e$, $\Omega$ and $\epsilon_0$ are 
the charge of the electron, volume of the cell and the permittivity of the 
free space, respectively. The operator $\hat{\mathbf{n}}$ represents the 
direction of the applied electric field. The real component of the 
dielectric constant can be derived from the imaginary component using 
the Kramers-Kronig relation \cite{gajdo06,Kreibig70}
\begin{equation}
\epsilon_1(\omega) = 1 + \frac{2}{\pi} P \int_0^{\infty} 
\frac{\epsilon_2(\omega')\omega' d\omega'}
{\omega'^2 - \omega^2 - i\eta},
\end{equation} 
where P is the principal value and $\eta$ is an infinitesimal broadening 
associated with the adiabatic switching of the dielectric perturbation.

The $\epsilon_1 (\omega)$ and $\epsilon_2 (\omega)$ for NBT and Ni-NBT from our calculations 
along with the available experimental data are shown in Fig. \ref{opt}.
For NBT, as discernible from panels (a) and (b), our calculated 
real and imaginary components of dielectric function are in good agreement 
with the experiment \cite{Bohdan-2009}. The slight deviation could be attributed 
to the temperature effects in experiment, as the reported experimental data are 
at room temperature.
Inspecting the real component more closely, the value of static dielectric 
constant, $\epsilon(0)$, is 6.5. This is consistent with the value reported 
in previous theory calculation \cite{Chandan-2023} and experiment \cite{Bohdan-2009}.
This relatively higher value of $\epsilon(0)$ suggests NBT as a potential 
candidate for light-harvesting applications \cite{Sudha-2018}.
The other important characteristic of real spectrum is, negative values at 
higher energies. This trend is consistent with the previously reported 
theoretical data \cite{Zeng-2010}, and suggests NBT as a potential candidate 
for plasmonic applications \cite{Shabani-2021}.

\begin{figure*}
	\centering\includegraphics[scale=0.5, angle=-90]{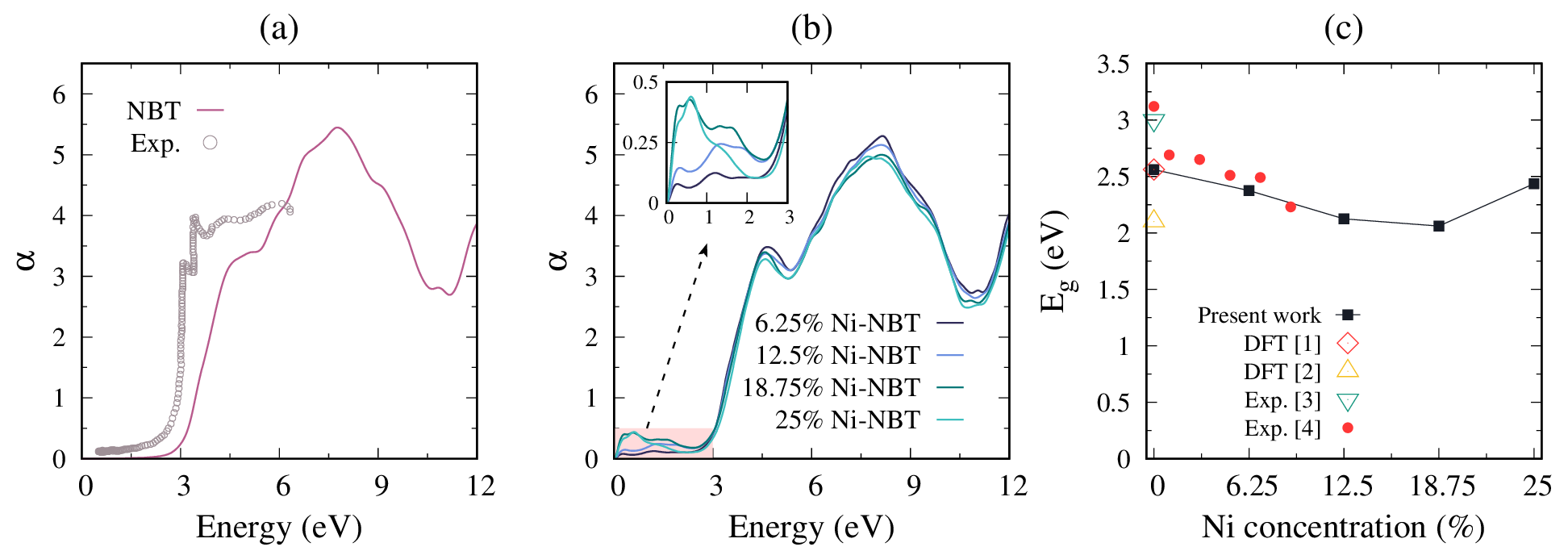}
	\caption{The absorption coefficient of NBT and Ni-substituted NBT
		(panels (a) and (b), respectively) and optical band gap as function 
		of Ni-concentrations (panel (c)).  Refs. [a], [b], [c] and
		[d] correspond to the works \cite{Chandan-2023}, \cite{Zeng-2010}, 
		\cite{Bohdan-2009} and \cite{Dung-2020}, respectively.} 
	\label{abs}
\end{figure*}

Examining the imaginary component more closely, we observe one preeminent 
peak at $\sim 4.2$ eV and four low-intensity secondary peaks at $\sim 5.7$, 
$\sim 6.4$, $\sim 7.7$, and $\sim 9.0$ eV energies. The primary peak originates 
from the interband transitions from O-$2p$ to Ti-$3d$ and Bi-$6p$ states. The 
secondary peaks, however, embed major contributions from the O-$2p$ to Na-$2s/2p$ 
transitions. Our calculated $\epsilon_2 (\omega)$ spectrum is in qualitative 
agreement with the reported theoretical data \cite{Zeng-2010}. The onset of 
$\epsilon_2 (\omega)$ spectrum suggests the optical band gap of NBT as 
$\approx2.57$ eV, which is consistent with the direct electronic bandgap 
discussed in previous section.

Considering the case of Ni-NBT, as discernible from the panels (c) and (d),
$\epsilon_1 (\omega)$ and $\epsilon_2 (\omega)$ show a similar trend as NBT at 
higher energies (above 2.5 eV). And, as can be inferred from Fig. \ref{fig4}(d), 
the reason for the observed peaks is attributed to the same interband 
transitions, O-$2p$ to Ti-$3d$ and Bi-$6p$. In the low energy regions (below 0.5 eV),
however, we observe a diverging nature of $\epsilon_1 (\omega)$. The reason for this 
could be attributed to the half-metallic nature of Ni-NBT. The value of static 
dielectric constant is observed to increase with Ni concentration, and the highest 
value of $\approx 61$ is obtained for 25$\%$ concentration. Consistent with the 
trend of $\epsilon_1 (\omega)$, $\epsilon_2 (\omega)$ shows sharp peaks below 
0.5 eV, with increasing amplitudes with Ni-concentration.

To get more insight and compare with experimental observations, we have examined 
the absorption coefficient, $\alpha$, for NBT and Ni-NBT.  
In addition, we have extracted the optical bandgap, $E_g$, for minority spin 
channel of Ni-NBT at different concentrations using the Tauc's plot \cite{Tauc-1966}. 
The data from this are shown in Fig. \ref{abs}. 
As discernible from the panel (a) of the figure, for NBT, our simulation is in 
good agreement with the experiment, with a slight shift in the onset of the peak.  
The reason for this shift could be attributed to the temperature effects in 
experimental data. For Ni-NBT, on contrary, we observe nonzero peaks in 
the IR region, which is consistent with the electronic structure data suggesting 
the half-metallic nature Ni-NBT. Panel (c) shows the optical bandgap for 
different Ni-concentrations. Consistent with the trend in experiment 
our computed $E_g$ decreases with Ni concentration. This could be 
explained in terms of inverse relationship between static dielectric constant 
and $E_g$ using the Penn model \cite{Penn-1962}. The increased $E_g$ for 25\% 
concentration could be attributed the decrease in $\epsilon_1 (0)$ to 61 from 81 for 18\%.  
Our computed $E_g$ for NBT is close to the previous calculations \cite{Chandan-2023}. 
The experimental values, Refs. \cite{Bohdan-2009, Dung-2020},
are however on the higher side as they are at finite temperatures.   

\subsection{Magnetic Properties}

\begin{figure}[!ht]
  \centering\includegraphics[scale=0.38, angle=0]{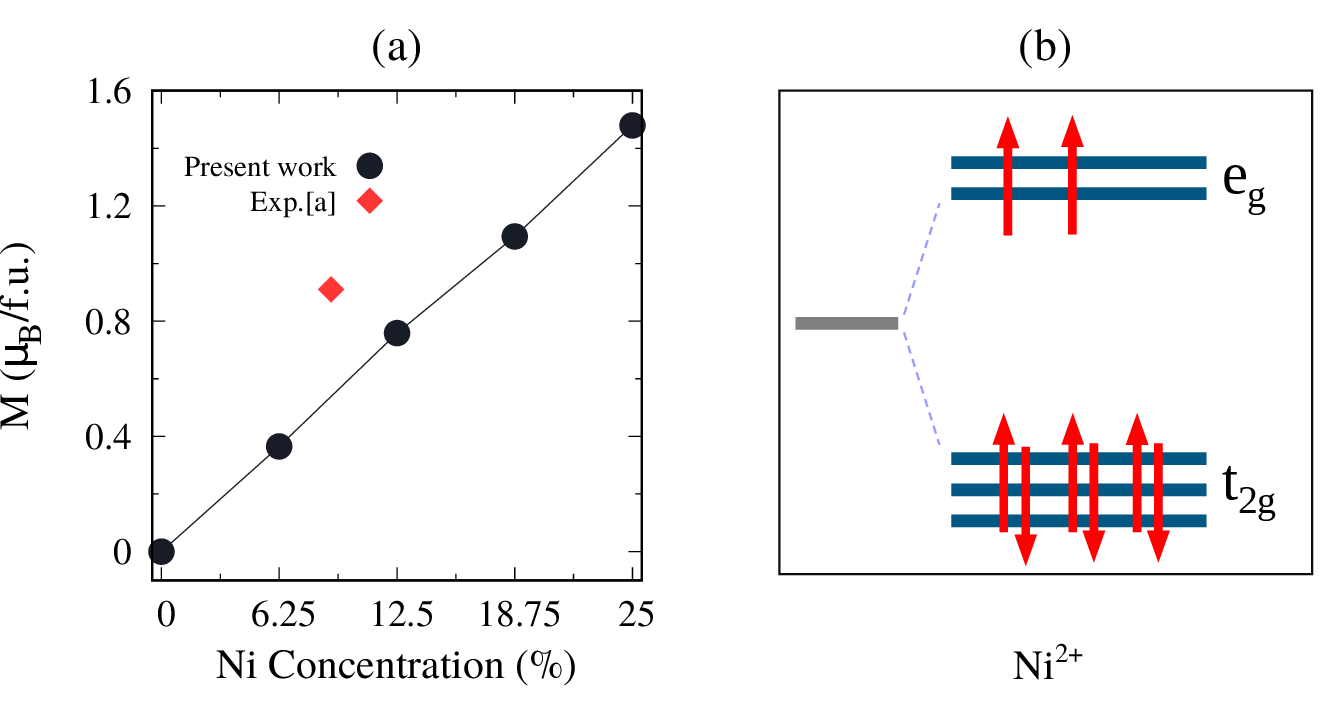}
  \caption{(a) Magentic moment as function of Ni concentration in Ni-NBT.  
	(b) Octahedral filling of electrons in $d$-orbitals. 
	The experimental value in panel (a) is from the Ref. \cite{Dung-2020}.} 
\label{mag}
\end{figure}

Next, as a probe to magnetic degrees of freedom introduced in the system, we 
examined the magnetic moments of NBT and Ni-NBT. And, to find the actual ground 
state magnetic configuration of the system, we probed both ferromagnetic (FM) 
and antiferromagnetic (AFM) orientations of magnetic moments. From our calculations, 
we find FM phase as the actual ground state for all concentrations Ni. This is 
evident from the relative energies of FM and AFM phases given in Table \ref{mag_tab} 
for 25\% of Ni, where the AFM energy is observed to be 
larger $\approx$ 25 meV.
In Fig. \ref{mag}, we have shown total magnetic moment as function of 
Ni-concentration.  Consistent with literature, and as to be expected due 
to pure ferroelectric nature, we obtained a zero magnetic moment for NBT.  
For Ni-NBT, however, we observed a trend of increasing (nonzero) magnetic moments 
as function of Ni-concentrations. The maximum magnetic moment observed 
is $~1.48$ $\mu_B$/f.u., for the highest concentration of 25\%.
The increase in the magnetic moment with concentration could be attributed
to the increasing ferromagnetic exchange between neighboring Ni ions at 
higher concentrations. Our calculated magnetic moment, 0.76 $\mu_B$/f.u., 
for concentration 12.5\% is in good agreement with the experimental value 0.91, 
reported for 9 \% of Ni \cite{Dung-2020}.

\begin{table}[ht!]
\caption{Relative energies of FM and AFM configurations, and 
	the atom resolved magnetic moments (in $\mu_B$/atom) for 25Ni-NBT.
	Contribution from the orbital magnetic moment is given in the 
	parenthesis.}
\centering
\begin{ruledtabular}
        \begin{tabular}{ll}
                E$_{\rm{FM}}$ (meV/f.u.)    & 0  \\
                E$_{\rm{AFM}}$ (meV/f.u.)  & 24.9 \\
                \hline
                $m_{\rm s}^{\rm Ni}$ ($m_{\rm o}^{\rm Ni}$)  & 1.539 (0.02)     \\
                $m_{\rm s}^{\rm O}$ ($m_{\rm o}^{\rm O}$)    & -0.122 (-0.006)  \\
                $m_{\rm s}^{\rm Ti}$ ($m_{\rm o}^{\rm Ti}$)  &  0.036 (0.0)     \\
                $m_{\rm s}^{\rm Bi}$ ($m_{\rm o}^{\rm Bi}$)  & 0.012 (-0.001)   \\
                $m_{\rm s}^{\rm Na}$ ($m_{\rm o}^{\rm Na}$)  & 0.001 (0.0)      \\
                \hline
                m{$^{\rm Tot}$}($\mu_B$/f.u.)   &  1.479
                \label{mag_tab}
        \end{tabular}
\end{ruledtabular}
\end{table}

To get more insight into the origin of nonzero magnetic moments in Ni-NBT, we 
examined separate contributions from each ion. The data from this is 
tabulated in Table \ref{mag_tab} for the highest concentration of 25\%. 
The values listed in the parenthesis are the contributions from orbital 
magentic moment.
As to be expected, Ni 
contributes dominantly, with $\approx$ 105\% of the total magnetic moment.  
The spin magentic moment originates from the unpaired $3d$-electrons in 
$e_g$ states (panel (b)). The obtained value of spin magentic moment, 1.54 $\mu_B$/atom, 
is however smaller than the expected theory value of 2.83 $\mu_B$/atom. 
The reason for this could be attributed to the strong hybridization 
between O $2p$ and Ni $3d$ orbitals.
Ni is also observed to display a small orbital magnetic moment of 
0.02 $\mu_B$/atom parallel to the spin contribution through SoC. 
The second dominant contribution is from the O ions. They contribute 
$\approx$ -9\% of the total value. The opposite contribution from O 
leads to a decrease in the total magentic moment. Like Ni, O also has 
a small parallel contribution from the orbital magentic moment. 
Among the other ions, Bi contributes about 2\% of total value, whereas 
Na and Ti has less than 1\% contributions.

\subsection{Magneto-optical Properties}

\begin{figure*}[!ht]
	\centering\includegraphics[scale=0.5, angle=0]{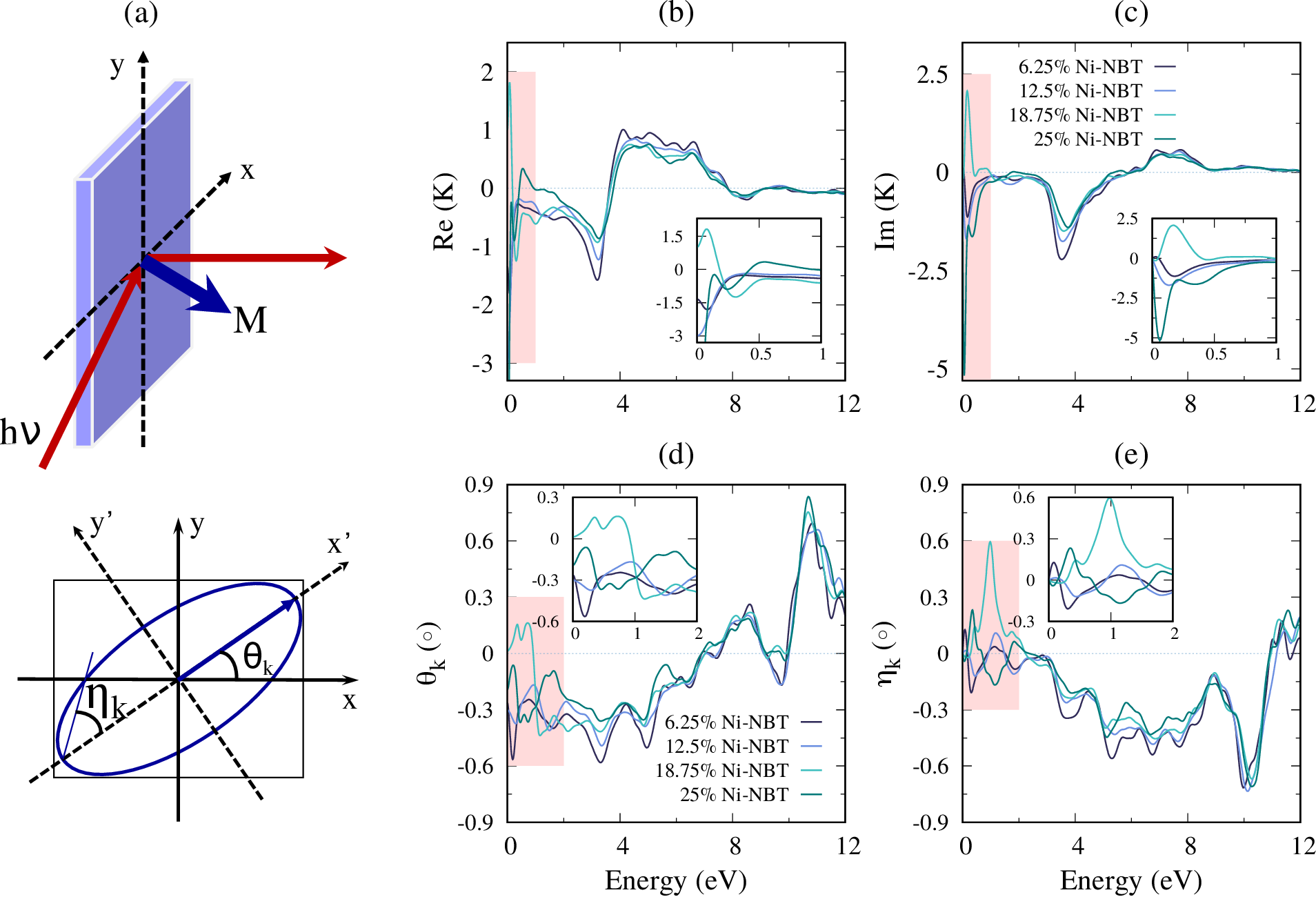}
	\caption{Real and imaginary components of magneto-optical permittivity
		(panels (a) and (b), respectively), Kerr rotation angle (panel (c))
		and Kerr ellipticity (panel (d)) as function of energy.}
	\label{moke}
\end{figure*}

The presence of magnetic degrees of freedom in Ni-NBT leads to an anisotropy 
in the dielectric tensor due to the breaking of time-reversal symmetry. 
The dielectric tensor for a magnetized material could be written as
\begin{equation}
      \epsilon_{ij} = \epsilon^{(0)}_{ij} + \epsilon^{(1)}_{ij}, 
\end{equation}
where $\epsilon^{(0)}_{ij}$ is dielectric tensor in absence of 
magnetization and $\epsilon^{(1)}_{ij}$ represents the contribution due 
nonzero magnetization. Within linear in magnetization $M$, $\epsilon^{(1)}_{ij}$ 
could be expressed as $\epsilon^{(1)}_{ij} = K_{ijk} M_k$, where $K$ is the 
magneto-optical coefficient.
To examine the magneto-optical properties of Ni-NBT, we computed MOKE 
spectra in the polar configuration (Fig. \ref{moke}(a)), where both 
the incident linearly-polarized wave and magnetization are considered 
perpendicular to the surface. The polar configuration is one of the most 
common setups used to trace the magneto-optical properties experimentally.  
The Kerr rotation angle, $\theta_k$, and Kerr ellipticity, $\eta_k$, can be 
extracted from the diagonal and off-diagonal dielectric response 
as \cite{picozzi06, sangalli12}
\begin{equation}
	\theta_k + i \eta_k = - \frac{K}{\sqrt{\epsilon^{(0)}} 
				(1 - \epsilon^{(0)})},
	\label{theeta}
\end{equation}
where $\epsilon^{(0)}$ is the diagonal component of the dielectric tensor 
in the absence of magnetization. Separating real and imaginary components 
in Eq. (\ref{theeta}), we can derive 
\begin{eqnarray}
	\theta_k &=& - \frac{[K_1^2 + K_2^2]^{1/2}}
	{[\epsilon_1^2 + \epsilon_2^2]^{1/4} 
	[(1-\epsilon_1)^2 + \epsilon_2^2]^{1/2}} \cos \Theta  {\;\;\; \rm and} \\
	\eta_k &=& - \frac{[K_1^2 + K_2^2]^{1/2}}
	{[\epsilon_1^2 + \epsilon_2^2]^{1/4} 
	[(1-\epsilon_1)^2 + \epsilon_2^2]^{1/2}} \sin \Theta,
\end{eqnarray}
where $\Theta = \tan^{-1}(\frac{K_2}{K_1}) 
- \frac{1}{2} \tan^{-1}(\frac{\epsilon_2}{\epsilon_1}) 
- \tan^{-1}(\frac{- \epsilon_2}{1 - \epsilon_1})$. 
The calculated real and imaginary parts of K and complex Kerr rotation 
angles for different concentrations of Ni-NBT are shown in Fig \ref{moke}.

As discernible from the panel (d), Kerr rotation shows the same qualitative 
behavior for all concentrations at higher energies, above $\sim 2$ eV. There are
substantial peaks in both negative and positive $y$-axes, which are signatures
of clockwise and anticlockwise polarizations, respectively, in the material.
In the low energy range,  below  $\sim 2$ eV, however, we observed a mix trend 
for $\theta_k$ at different concentrations.
In the negative $y-$axis, the most significant peak of amplitude 0.58$^\circ$ is
observed at 3.3 eV for 6.5\% concentration. The amplitude of the peaks is 
observed to decrease with  Ni-concentration. In the positive $y$-axis, however, 
a $\theta_k$ reaching up to 0.7$^\circ$ around 10.8 eV is observed for 25\% 
concentration. Unlike the trend for negative $\theta_k$, the peak amplitude
for positive $\theta_k$ increases with Ni-concentrations.

Fig. \ref{moke}(e) shows the Kerr ellipticity data as function of energy. 
Like the trend of $\theta_k$, all concentrations show a similar qualitative 
behavior at higher energies, whereas a mix trend for amplitudes is observed in
the low energy range. We observed a significant peak of amplitude 
0.72$^\circ$ around 10 eV energy in the negative $y$-axis for 6.25\% concentration. 
Unlike $\theta_k$, there is not much variation in the amplitude of this peak 
with Ni-concentrations. Consistent with the trend of $\theta_k$, apart from this 
primary peak, we also observe few secondary peaks of amplitudes 0.33$^\circ$, 
0.56$^\circ$, and 0.54$^\circ$ at 4.0, 5.25 and 6.7 eV, respectively. 
The significant Kerr signals obtained from our simulations for Ni-NBT suggest 
it as a potential candidate for magneto-optical applications.  

\section{Conclusions}

In conclusion, with the help of density functional theory based first-principles
calculations, we examined the effect of transition metal substitution on
electronic, ferroelectric, magnetic, optical and magneto optical
properties of NBT. In agreement with literature, our simulations on electronic
properties show NBT as a direct band semiconductor. Our computed bandgap
2.56 eV is within the range of previous theory calculations and experiments.
For transition metal substituted-NBT, we observed an emergence of half-metallic
ferromagnetism in the system. Our simulation show, while minority spin exhibits
a wide bandgap, there are nonzero states at Fermi energy for majority spin.
The reason for this could be attributed to the shift in the energy levels
of majority spin states due to hybridization between O $2p$ and Ni $3d$
states. This asymmetry in the two spin channels lead to an emergence of
nonzero permanent magentic moment in the material. We obtained a magnetic
moment of 1.5 $\mu_{\rm B}/{\rm f.u.}$ for 20\% of Ni concentration.
For optical properties of NBT, our simulation results are consistent
with the available experimental and other theory results.
For Ni-NBT, however, we observed a diverging nature of static dielectric
constant in the infrared region, which could be attributed the metallic
nature of the material. Our data on MOKE show significant values of
Kerr angles in Ni-NBT, which suggests transition metal substituted-NBT
as potential candidates for magneto-optical applications.

\section*{Acknowledgments}

The authors wish to thank Ravi Kumar, Mohd Zeeshan and Indranil Mal 
for useful discussions. C. K. V. acknowledges the funding support from 
Council of Scientific \& Industrial 
Research, India (Grant No. $09$/$086$($1297$)/$2017$-EMR-I).
B. K. M. acknowledges the funding support from SERB, DST (CRG/2022/003845).
The results presented in the paper are based on the computations using the
High Performance Computing cluster, Padum, at the Indian Institute of
Technology Delhi, New Delhi


\bibliography{reference}
\end{document}